**Comments regarding "Transonic dislocation propagation in diamond" by Katagiri, et al. (*Science* 382, 69-72, 2023)**

J. A. Hawreliak,[1] J. M. Winey,[1] Surinder. M. Sharma,[2] and Y. M. Gupta[1,3]

[1] *Institute for Shock Physics, Washington State University, Pullman, Washington 99164, USA*
[2] *Physics Group, Bhabha Atomic Research Centre, Mumbai 400 085, India*
[3] *Department of Physics and Astronomy, Washington State University, Pullman, Washington 99164, USA*

We have carefully examined the above-referenced paper and find the claims of stacking fault formation and transonic dislocation propagation in diamond to be not valid. Additionally, it is quite puzzling that 14 authors on this paper are also co-authors on another recent paper that directly conflicts with the dislocation claims in the *Science* paper.

I. Introduction

In their Research Article in *Science*, Katagiri et al. (*1*) presented *in situ* x-ray imaging results for laser-shocked diamond single crystals. As shown in Fig. 1, the drive laser (260 μm spot size) was incident on a 50 μm thick polypropylene ablator affixed to one face (500 μm x 500 μm) of the diamond sample, which was 200-300 μm thick. The XFEL beam was incident on the diamond sample normal to the drive laser, and the transmitted x-rays were detected using a LiF crystal. Since each experiment provides one image, the imaging results obtained at various times represent a compilation of images from different experiments. Because wave profile measurements were not made in these experiments, reproducibility of the propagating shock wave in the different experiments was not established in Ref. 1.

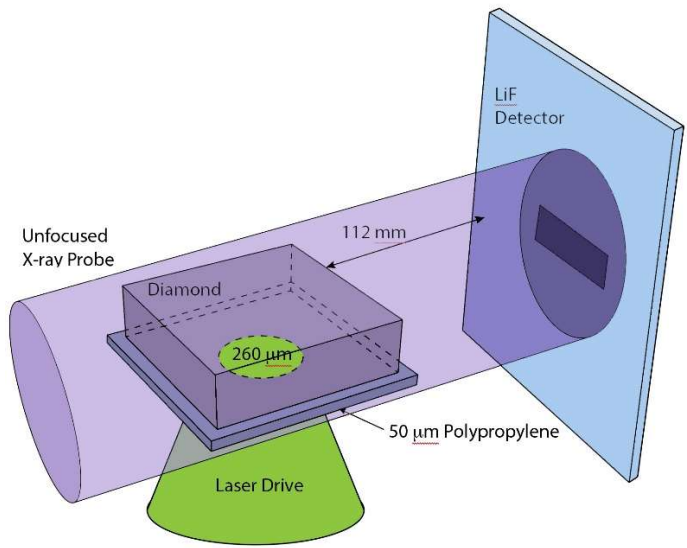

Figure 1 – Schematic representation of the experimental configuration in Ref. 1.

The x-ray images for shock compressed diamond in Katagiri, et al. (see Figs. 2B and 2D in Ref. 1) show a two-wave structure, with long (many micrometers) striations appearing in the images behind the second wave front. The first and second waves were denoted as elastic and plastic shock waves by Katagiri, et al., and the striation features were interpreted as stacking faults (*1*). Furthermore, the authors claimed that the growth of the stacking faults in their image data implies the transonic propagation of partial dislocations in diamond, where "transonic" indicates propagation speeds greater than the transverse sound speed and less than the longitudinal sound speed. However, we note that their claims are based solely on the appearance of striations in their images and the authors provide no justification for identifying these striations as stacking faults.

The claims of stacking fault formation and transonic dislocation motion (*1*) are in direct conflict with the brittle fracture claim of Makarov, et al. (*2*), who conducted x-ray imaging experiments on laser-shocked diamond single crystals using the same experimental approach and the same XFEL facility (as Ref. 1) and published their results a few weeks earlier. Similar to Katagiri, et al. (*1*), Makarov, et al. (*2*) observed a two-wave structure in their x-ray images. However, based partly on numerical simulations of their laser-shock experiments, they interpreted their results in terms of a brittle fracture response, not dislocation-based plasticity. The conflicting interpretations and claims published by Katagiri, et al. (*1*) and Makarov, et al. (*2*) are quite puzzling, since the two papers were published within weeks of each other and 14 authors on the Katagiri, et al. paper (including K. Katagiri) are co-authors on the Makarov, et al. paper. The conflicting claims in these two papers raise the following questions: how can these 14 authors support conflicting claims regarding inelastic deformation in diamond and which of the two claims should the reader believe?

The remainder of our report shows that brittle fracture is a more realistic mechanism for inelastic deformation in shock compressed diamond and we point out numerous substantive errors in the Katagiri, et al. paper (*1*).

II. Inelastic deformation in shock compressed diamond

Shock wave compression results in significant shear stress build up in solids. When the shear stresses exceed the material's elastic limit, inelastic deformation occurs in order to relax the shear stresses. In metals, inelastic deformation typically occurs through the motion and generation of dislocations and is commonly denoted as plasticity (*3*); in strong covalent solids, such as diamond, inelastic deformation typically occurs via brittle failure mechanisms such as cracking (*4*).

The claims of stacking fault formation and transonic dislocation motion by Katagiri, et al. (*1*) assume a metal-like elastic-plastic response. However, the high stacking fault energy of diamond (~280 mJ/m$^2$ (*5*)) would preclude the formation and growth of the large (many micrometers) stacking faults required for the claims in Ref. 1; in fact, the stacking fault widths measured in diamond are orders of magnitude smaller (<10 nm) (*5*). Instead of an elastic-plastic response, the brittle fracture response – incorporated by Makarov, et al. based on similar images (*2*) – is a more realistic inelastic deformation mechanism for diamond and is in agreement with previously published results for shock compressed diamond single crystals (*6,7*). Both laser shock (*6*) and plate impact (*7*) experiments clearly showed that shock-induced inelastic deformation in diamond is incompatible with an elastic-plastic response and requires, instead, a brittle fracture description. In addition, the {111} planes – identified by Katagiri, et al. with the observed striations in their images – are well known as the preferred diamond cleavage planes (*8*).

Although brittle fracture, and not dislocation motion, governs inelastic deformation in shocked diamond single crystals, detailed description of a time-dependent brittle fracture model for shocked diamond single crystals remains an important need. Finally, as pointed out in the next section, there is no justification for relating the observed striations (*1*) to dislocation motion.

III. Assessment of experiments reported by Katagiri, et al. (*1*)

A. Shock wave loading

In the experiments by Katagiri, et al. (*1*), shock wave amplitudes were not measured. Knowledge of the input stress history into the diamond samples or a measurement of the propagating wave – common practice in laser-shock experiments (*9*) – is essential to analyze and interpret the x-ray data. Below, we describe our approach to determining the input stress history (into the diamond) for the experiments reported in Ref. 1 and the resulting conclusions regarding the x-ray measurements.

To determine the stress history at the polypropylene ablator/diamond interface in the experiments by Katagiri et al. (*1*), we performed two simulations using the radiation hydrodynamics code HYADES (*10*). The first simulation used an idealized flat top laser pulse with the laser energy, pulse length, and focal spot values published by Katagiri et al. For this simulation, the laser temporal profile was modelled as a 5ns flat top pulse with a rise and fall time of 100ps, delivering 17 J in a 260 μm diameter spot. This gives a focal spot intensity of I = $6.16 \times 10^{12}$ W/cm$^2$ at 532 nm. The simulated laser intensity as a function of time is shown in Fig. 2 a) in blue. The second simulation used the temporal laser profile provided by Makarov at al. (*2*) in Fig. S1. We scaled the intensity given by Makarov et al. to match the total energy of 17 J from Katagiri, et al., while maintaining the temporal profile published by Makarov, et al. The simulated intensity as a function of time for this scaled laser pulse is shown in Fig. 2 a) in red.

Figure 2 b) shows an x-t plot of the longitudinal stress response in the ablator for the ideal flat top pulse. Figure 2 c) shows that there are relatively small differences in the stress history at the diamond-ablator interface between the two pulse shapes. As shown in Fig. 2 b), at the laser onset, an initial shock wave of 190 GPa is generated in the ablator. The laser-plasma interaction of the plasma plume in the ablator decreases the ablation stress over time (*11*), and the shock wave reaching the diamond has a stress of ~150 GPa. Due to the impedance mismatch between the ablator and diamond, a >300 GPa shock wave propagates into the diamond and a >300 GPa shock is reflected back in the ablator. The stress history at the ablator-diamond interface, shown in Fig. 2 c), shows a gradual reduction in stress until the stress is rapidly released to <100 GPa due to reverberations of the shock in the ablator, Fig 2 b). The rapid drop in the stress occurs only 2.2 ns after the initial shock. At later times, additional stress release occurs due to the arrival of the release wave from the end of the laser pulse.

The calculated stress histories shown in Fig. 2 c) – which are consistent with that from Makarov, et al. (*2*) – contradict the stress estimates that were provided by Katagiri, et al. (*1*). In particular, the >300 GPa calculated initial stress is much larger than the 184 GPa and 92 GPa stress for [100] diamond and [110] diamond, respectively, estimated in Ref. 1.

Furthermore, the initial stress reduction behind the shock wave followed by the rapid stress release to <100 GPa (after 2.2 ns) results in a rapidly travelling unloading wave into the diamond; the initial release travels at the longitudinal sound speed in the shocked state. Because of the faster propagating release wave, the much slower plastic wave speeds reported by Katagiri, et al. (*1*) are not meaningful. In fact, the wave analysis approach used by Katagiri, et al. is not applicable due to the input stress loading imparted to their diamond samples. As such, inferences from their x-ray images regarding the shocked states in diamond are not valid.

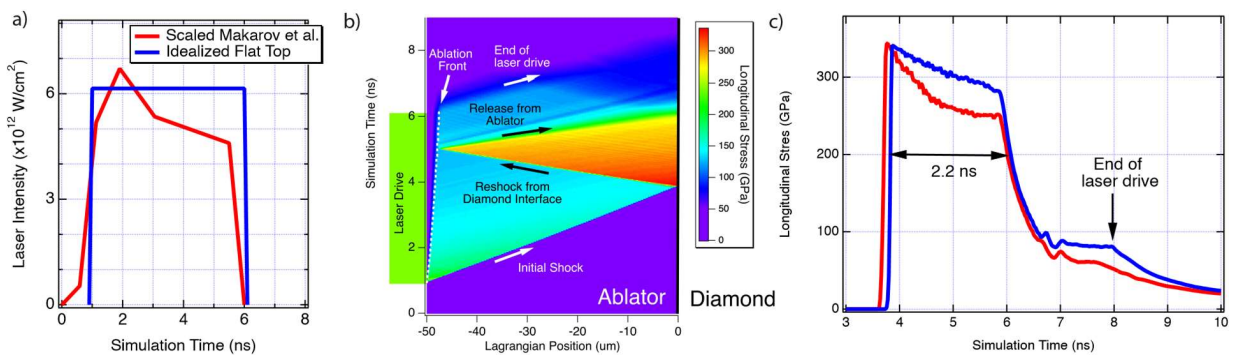

Figure 2: a) Simulated laser profiles. The idealized flat top is shown in blue and an energy scaled pulse from the laser profile published in Makarov et al. (*2*) is shown in red. b) x-t plot of the stress in the ablator due to the idealized flat top pulse. The green region denotes the time the drive laser is on. c) Stress at the ablator/diamond interface as a function of time.

B. X-ray imaging measurements

Katagiri et al. (*1*) claim that the observed linear features oriented parallel to the {111} planes in their measured images are due to x-ray scattering from stacking faults, assumed to form on the {111} planes in shock compressed diamond. They also state that their experimental method is distinct from phase contrast imaging. However, we note that their experimental setup is in fact the standard configuration for phase contrast imaging and the images they report show very clear phase contrast features. In addition, the experimental configuration used in Ref. 1 is essentially identical to that used by Makarov, et al. (*2*), who note that this approach is phase contrast imaging.

Because stacking faults do not contribute to density variations in a material, they do not give rise to features in phase contrast images. It has been well documented in theory, simulations, and experiments that x-ray scattering from atoms across a stacking fault contributes to a phase shift of the x-ray diffraction signal but does not generate its own scattering signal (*12,13*). Experimentally, stacking fault densities are typically determined using x-ray diffraction from multiple lattice planes and measuring the deviation in the Bragg scattering angle from the expected scattering angle (*14,15*). Imaging stacking faults in diamond requires a diffraction contrast topographic configuration (*16*); such measurements, if made, would also necessitate substantial modeling and simulations of images. Direct imaging of stacking faults is very weak and would not be observable using the imaging approach in Ref. 1.

Similarly, measurements of dislocations in materials have been performed using diffractive techniques (*17-19*), not imaging techniques. The scattering from dislocations is so weak that it would be impossible to see in the transmitted beam used by Katagiri et al. (*1*).

As discussed in section II, brittle fracture is the relevant inelastic deformation mechanism in shock compressed diamond single crystals (*6,7*). Phase contrast imaging is well suited for resolving cracks – in contrast to stacking faults and dislocations – because of the density variations inherent to cracking. Numerical simulations by Makarov et al. (*2*), using their brittle fracture model for diamond, show extensive cracking, especially around the periphery of the laser drive spot where the loading histories and stress states are complex.

To determine if cracking can account for the x-ray image features reported by Katagiri, et al. (*1*), we carried out phase contrast imaging simulations using the model shown in Fig. 3 a). This model consists of a single conically shaped crack (shown in red) where the facets of the crack are along {111} planes; the hemispherical elastic wave front is shown in blue. Figures 3 b) and c) show the simulated phase contrast images when the x-ray beam is oriented along the [011] and [001] diamond directions, respectively; the [011] orientation is that used by Katagiri, et al. (*1*). Figure 3 b) shows features consistent with those reported in Ref. 1: linear features following the {111} planes, with no significant features in the center of the image and the density of features increasing toward the lateral boundaries. Figure 3 c) shows that the features are less observable when viewed along [001], consistent with results reported by Katagiri, et al. in their supplemental material (*1*).

The discussion and simulations presented in this section make a strong case that the features in the images reported by Katagiri, et al. are not due to scattering from stacking faults or dislocations. Instead, they are due to the enhanced contrast from cracks along {111} planes in shock compressed diamond.

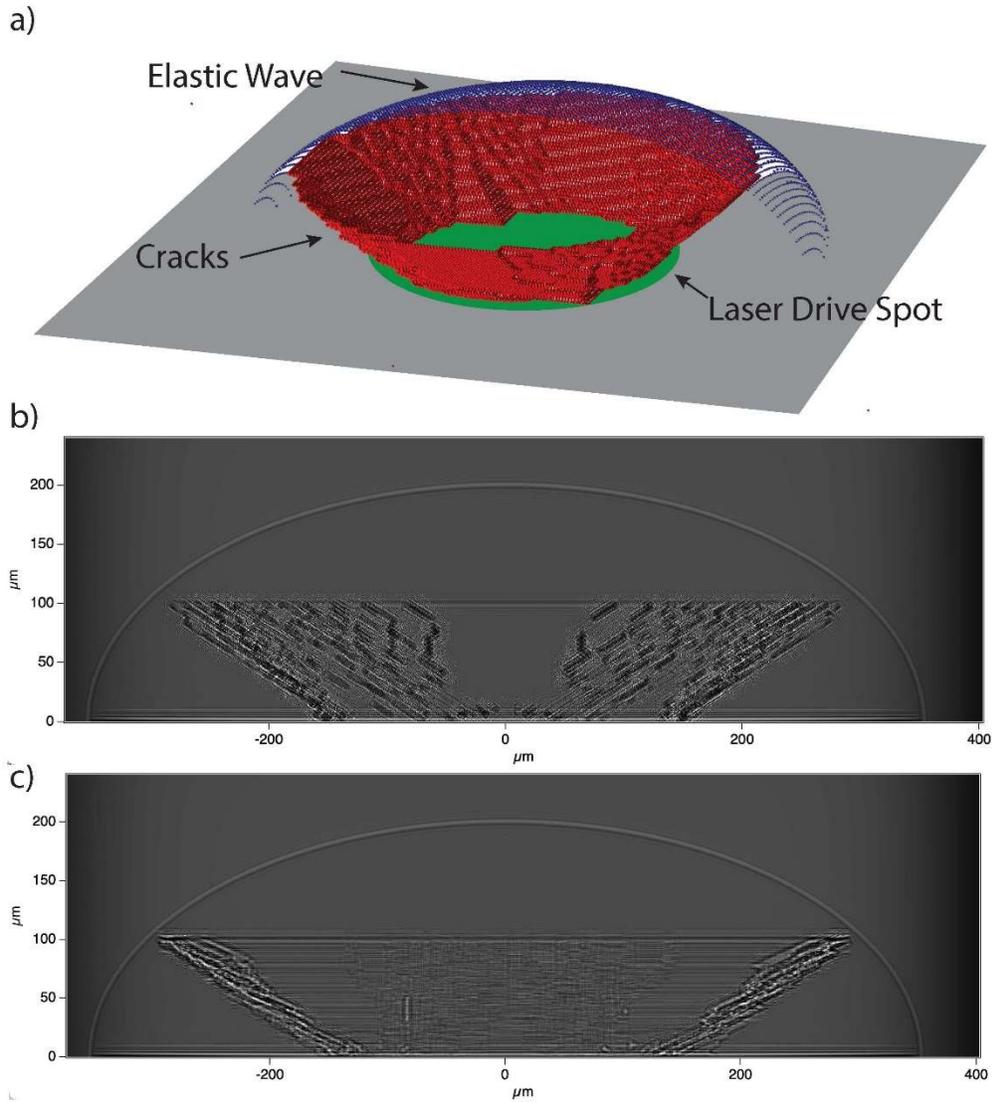

Figure 3: a) Crack model and elastic wave configuration. b) and c) Simulated phase contrast images using the experimental setup of Katagiri, et al. (*1*) for the [011] and [001] imaging directions, respectively.

IV. Concluding remarks

    The analysis and discussion presented in this report show that the claims of stacking fault formation and transonic dislocation propagation by Katagiri, et al. (*1*) are simply not valid. Because of a number of factors – x-ray imaging technique not being sensitive to stacking faults, lack of wave profile measurements, substantive errors in wave propagation analysis, and 14 authors (including the lead author) supporting conflicting claims regarding inelastic deformation in shocked diamond single crystals – the credibility of the work in Ref. 1 is questionable.